\documentclass{elsart}

\usepackage{epsfig}

\newcommand{\be}{\begin{equation}}
\newcommand{\ee}{\end{equation}}
\newcommand{\bea}{\begin{eqnarray}}
\newcommand{\eea}{\end{eqnarray}}

\begin{document}

\begin{frontmatter}
\title{Charmonium dynamics in nucleus-nucleus collisions at SPS and FAIR energies}
\author[FIAS]{O.~Linnyk,\corauthref{cor1}}
\ead{linnyk@fias.uni-frankfurt.de} \corauth[cor1]{corresponding
author}
\author[FIAS]{E. L. Bratkovskaya,}
\author[unig]{W.~Cassing,}
\author[FIAS,unif]{H. St\"ocker}
\address[FIAS]{
Frankfurt Institute for Advanced Studies,
 Johann Wolfgang Goethe University,
 Max-von-Laue-Str. 1,
 60438 Frankfurt am Main,
 Germany}
\address[unif]{
Institut f\"ur Theoretische Physik, %
 Johann Wolfgang Goethe University,
 Max-von-Laue-Str. 1,
 60438 Frankfurt am Main,
 Germany}
\address[unig]{Institut f\"ur Theoretische Physik, %
  Universit\"at Giessen, %
  D--35392 Giessen, %
  Germany}

\begin{abstract}
Charmonium production and suppression in In+In and Pb+Pb reactions
at SPS energies is investigated with  the HSD transport approach
within the `hadronic comover model' as well as the `QGP threshold
scenario'. The results of the transport calculations for $J/\Psi$
suppression and the $\Psi^\prime$ to $J/\Psi$ ratio are compared
with the recent data of the NA50 and NA60 Collaborations. We find
that the comover absorption model -- with a single parameter
$|M_0|^2$ for the matrix element squared for charmonium-meson
dissociation -- performs best with respect to all data sets. The
`threshold scenario' -- within different assumptions for the melting
energy densities -- yields a reasonable suppression for $J/\Psi$ but
fails in reproducing the $\Psi^\prime$ to $J/\Psi$ ratio for Pb+Pb
at 158 A$\cdot$GeV.  Predictions for Au+Au reactions are presented
for a bombarding energy of 25 A$\cdot$GeV in the different scenarios
which will allow for a clear distinction between the models from the
experimental side at the future FAIR facility.
\end{abstract}


\end{frontmatter}



\section{Introduction}

The dynamics of ultra-relativistic nucleus-nucleus collisions at
Super-Proton-Synchrotron (SPS) and Relativistic-Heavy-Ion-Collider
(RHIC) energies are of fundamental interest with respect to the
properties of hadronic/partonic systems at high energy densities.
Especially the formation of a quark-gluon plasma (QGP) and its
transition to interacting hadronic matter has motivated a large
community for almost three decades~\cite{QM01}. The $c, \bar{c}$
 quark degrees of freedom are  of particular interest in
context to the phase transition to the QGP, since $c\bar{c}$ meson
states might no longer be formed due to color screening
\cite{Satz,Satznew,Satzrev,KSatz}. However, more recent lattice QCD
(lQCD) calculations have shown that the $J/\Psi$ survives up to at
least 1.5 $T_c$ ($T_c \approx$ 170 MeV) such that the lowest
$c\bar{c}$ states remain bound up to rather high energy
density~\cite{KarschJP,HatsudaJP,Karsch2}. On the other hand the
$\chi_c$ and $\Psi^\prime$ appear to melt soon above $T_c$. It is
presently not clear if also the $D$ or $D^*$ mesons will survive at
temperatures $T > T_c$ but strong correlations between a light quark
(antiquark) and a charm antiquark (quark) are likely to persist
\cite{Rapp05}.

The standard approach to charmonium production in heavy-ion
collisions assumes that $c\bar{c}$ pairs  are created exclusively at
the initial stage of the reaction in primary nucleon-nucleon
collisions. At the very early stage color dipole states are expected
to be formed which experience i) absorption by interactions with
nucleons of the colliding nuclei ({\it cf.} Refs.
\cite{Kharzeev,Capella}). These $c\bar{c}$ states are assumed to be
absorbed in a `pre-resonance state' before the final hidden charm
mesons are formed.  This absorption -- denoted by `normal nuclear
suppression' -- is also present in p+A reactions and determined by a
dissociation cross section $\sigma_B$ $\sim$ 4 to 7 mb). Those
charmonia or `pre-resonance' states -- that survive normal nuclear
suppression -- furthermore suffer from ii)  a possible dissociation
in the deconfined medium at sufficiently high energy density and
iii) the interactions with secondary hadrons (comovers) formed in a
later stage of the nucleus-nucleus collision.

The geometrical Glauber model of Blaizot et al.~\cite{Blaizot}, as
well as the percolation model of Satz~\cite{Satzrev}, assumes that
the QGP suppression ii) sets in rather abruptly as soon as the
energy density exceeds a threshold value $\varepsilon_c$, which is a
free parameter. This version of the standard approach will be
referred to as the QGP `threshold scenario'. The latter model is
motivated by the idea that the charmonium dissociation rate is
drastically larger in a quark-gluon-plasma (QGP)  than in a hadronic
medium \cite{Satzrev} such that further comover absorption channels
might be neglected.

On the other hand, the extra suppression of charmonia in the high
density phase of nucleus-nucleus collisions at SPS energies
\cite{NA50aa,NA50b,NA50a,NA38,NA50O,NA60} has been attributed to
inelastic comover scattering ({\it cf.}
\cite{Capella,Cass99,Vogt99,Gersch,Cass00,Kahana,Spieles,Gerland}
and Refs. therein) assuming that the corresponding $J/\Psi$-hadron
cross sections are in the order of a few mb
\cite{Haglin,Konew,Ko,Sascha}. Theoretical estimates here differ by
more than an order of magnitude~\cite{Bernd} especially with respect
to $J/\Psi$-meson scattering such that the question of charmonium
suppression is still open. Additionally, alternative absorption
mechanisms -- such as gluon scattering on color dipole states --
might play a role as suggested in
Refs.~\cite{Kojpsi,Rappnew,Blaschke1,Blaschke2} and also lead to a
reduction of the final $J/\Psi$ formation in central nucleus-nucleus
collisions.

We recall that apart from absorption or dissociation channels for
charmonia also recombination channels such $D+ \bar{D} \rightarrow
J/\Psi$ + meson may play a role. A previous analysis within the HSD
transport approach~\cite{brat03} -- employing the comover absorption
model -- has demonstrated that the charmonium production from open
charm and anticharm mesons indeed becomes essential in central Au+Au
collisions at RHIC. This is in accordance with independent studies
in Refs.~\cite{Ko,Rappnew} and also with the data from PHENIX
\cite{PHENIX}. On the other hand, these backward channels --
relative to charmonium dissociation with comoving mesons -- have
been found to be practically negligible at the SPS energies of
interest here. Nevertheless, in our dynamical studies below we will
include the `backward' channels for completeness.

A couple of models have predicted $J/\Psi$ suppression in In+In
collisions as a function of centrality at 158 A$\cdot$GeV based on
the parameters fixed for Pb+Pb reactions at the same bombarding
energy. However, the predictions within the comover model and
`threshold scenario' from Refs.~\cite{NA60theo2,NA60theo1,NA60theo3}
have failed to describe the data with sufficient accuracy. This
might be either due to the missing dynamics of the nucleus-nucleus
collisions in these models or to inadequate physical assumptions
about the dissociation mechanism.

In the present work we extend the previous studies within the
comover model in Refs.~\cite{Cass01,brat03,brat04} and test the `QGP
threshold scenario' -- described in Section 2 -- in comparison to
the Pb+Pb data at 158 A$\cdot$GeV from NA50 as well the high
statistics data from NA60 for In+In collisions at the same
bombarding energy. The question we aim at solving in Section 3 is:
1) can any of the models be ruled out by the combined data sets and
2)  do the recent NA60 data provide a hint to QGP formation at the
top SPS energy? In Section 4 we, furthermore, will provide
predictions for the charmonium suppression in Au+Au collisions at 25
A$\cdot$GeV that will be measured at the future FAIR facility by the
CBM Collaboration.

\section{Brief description of charmonium channels in HSD}

\begin{figure}[!]
\psfig{figure=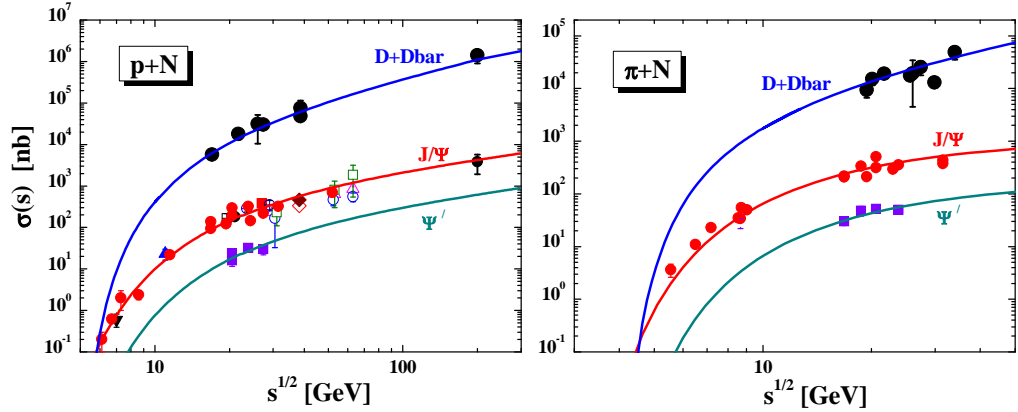,width=0.73\textwidth}
\caption{The cross section for $D+\bar D$, $J/\Psi$ and
$\Psi^\prime$ meson production in  $pN$ (left part) and $\pi N$
reactions (right part). The solid lines show the parametrisations
used in HSD, whereas the symbols stand for the experimental data
\protect\cite{NA16,NA27,E743,E653,E789,NA32,E769,WA92,E791,PHENIX_pp}.
The $J/\Psi$ cross sections include the decay from $\chi_c$ mesons.
} \label{xs_pp_pip}
\end{figure}

The microscopic Hadron-String-Dynamics (HSD) transport calculations
(employed here) provide the space-time geometry of the
nucleus-nucleus reaction and a rather reliable estimate for the
local energy densities achieved since the production of secondary
particles is described rather well from SIS to RHIC
energies~\cite{Weber}. In order to examine the dynamics of open
charm and charmonium degrees of freedom during the formation and
expansion phase of the highly excited system created in a
relativistic nucleus-nucleus collision within transport approaches,
one has to know the number of initially produced particles with $c$
or $\bar{c}$ quarks, i.e. $D, \bar{D}, D^*, \bar{D}^*, D_s,
\bar{D}_s, D_s^*, \bar{D}_s^*,$ $J/\Psi(1S), \Psi^\prime(2S),
\chi_c(1P)$. In this work we follow the previous studies in Refs.
\cite{Cass99,Cass00,brat03,Cass01} and fit  the total charmonium
cross sections ($i = \chi_c, J/\Psi, \Psi^\prime$) from $NN$
collisions as a function of the invariant energy $\sqrt{s}$ by the
expression
\begin{eqnarray}
\sigma_i^{NN}(s) = f_i \ a \ \left(1 - \frac{m_i}{\sqrt{s}}\right)^\alpha
\ \left(\frac{\sqrt{s}}{m_i}\right)^\beta \theta(\sqrt{s}-\sqrt{s_{0i}}),
 \label{fitj}
\end{eqnarray}
where $m_i$ denotes the mass of charmonium $i$ while
$\sqrt{s_{0i}}=m_i+2 m_N$ is the threshold in vacuum. The parameters
in (\ref{fitj}) have been fixed to describe the $J/\Psi$ and
$\Psi^\prime$ data up to the RHIC energy $\sqrt{s}=200$ GeV ({\it
cf.} Fig. \ref{xs_pp_pip}). We use $a=0.2$ mb, $\alpha$ = 10, $\beta
=0.775$. The parameters $f_i$ are fixed as $f_{\chi_c}=0.636, \
f_{J/\Psi}=0.581,\ f_{\Psi^\prime}=0.21$ in order to reproduce the
experimental ratio
$$\frac{B(\chi_{c1}\to J/\Psi)\sigma_{\chi_{c1}}
 +B(\chi_{c2}\to J/\Psi)\sigma_{\chi_{c2}}}
 {\sigma^{exp}_{J/\Psi}}=0.344\pm 0.031$$
measured in $pp$ and $\pi N$ reactions~\cite{E705_93,WA11_82} as
well as the averaged $pp$ and $pA$ ratio
$(B_{\mu\mu}(\Psi^\prime)\sigma_{\Psi^\prime})
 / (B_{\mu\mu}(J/\Psi)\sigma_{J/\Psi})\simeq 0.0165$ ({\it cf.} the compilation
of experimental data in Ref.~\cite{NA50_03}). Here the
experimentally measured $J/\Psi$ cross section includes the direct
$J/\Psi$ component $(\sigma_{J/\Psi})$ as well as the decays of
higher charmonium states $\chi_{c}, \Psi^\prime$, {\it i.e.}
\begin{eqnarray}
\sigma^{exp}_{J/\Psi}=\sigma_{J/\Psi}+B(\chi_{c}\to
J/\Psi)\sigma_{\chi_{c}} +B(\Psi^\prime\to
J/\Psi)\sigma_{\Psi^\prime}. \ \label{xsexp}\end{eqnarray} Note, we
do not distinguish the $\chi_{c1}(1P)$ and $\chi_{c2}(1P)$ states.
Instead, we use only the $\chi_{c1}(1P)$ state (which we denote as
$\chi_c$), however, with an increased  branching ratio for the decay
to $J/\Psi$ in order to include the contribution of $\chi_{c2}(1P)$,
{\it i.e.}  $B(\chi_{c}\to J/\Psi) = 0.54$. We adopt
$B(\Psi^\prime\to J/\Psi)=0.557$ from Ref.~\cite{PDG}.

In addition to primary hard $NN$ collisions the open charm mesons or
charmonia may also be generated by secondary meson-baryon ($mB$)
reactions. Here we include all secondary collisions of mesons with
baryons by assuming that the open charm cross section (from Section
2 of Ref.~\cite{Cass01}) only depends on the invariant energy
$\sqrt{s}$ and not on the explicit meson or baryon state.
Furthermore, we take into account all interactions of `formed'
mesons -- after a formation time of $\tau_F$ = 0.8 fm/c (in their
rest frame)~\cite{Geiss} -- with baryons or diquarks, respectively.
For the total charmonium cross sections from meson-baryon (or $\pi
N$) reactions we use the parametrization (in line with Ref.
\cite{Vogt99}):
\begin{eqnarray}
\sigma_i^{\pi N} (s) = f_i \ b \ \left(1 - \frac{m_i}{\sqrt{s}}\right)^\gamma
\label{fitpin}\end{eqnarray}
with $\gamma=7.3$ and $b=1.24$~mb, which describes the
existing experimental data at low $\sqrt{s}$ reasonably well
as seen from Fig. \ref{xs_pp_pip}.

Apart from the total cross sections, we also need the differential
distribution of the produced mesons in the transverse momentum $p_T$
and the rapidity $y$ (or Feynman $x_F$) from each individual
collision. We recall that $x_F = p_z/p_z^{max} \approx 2
p_z/\sqrt{s}$ with $p_z$ denoting the longitudinal momentum. For the
differential distribution in $x_F$ from $NN$ and $\pi N$ collisions
we use the ansatz from the E672/E706 Collaboration~\cite{E672}:
\begin{equation}
\frac{dN}{dx_F dp_T} \sim (1 - |x_F|)^c \ \exp(-b_{p_T} p_T),
\label{fit2}
\end{equation}
where $b_{p_T}=2.08$ GeV$^{-1}$ and $c= a/(1+b/\sqrt{s})$. The
parameters $a, b$ are chosen as $a_{NN}=13.5$, $b_{NN}=24.9$ for
$NN$ collisions and $a_{\pi N}=4.11$, $b_{\pi N}=10.2$ for $\pi N$
collisions as in~\cite{brat03,Cass01}.

The parametrizations of the total and differential cross sections
for open charm mesons from $pN$ and $\pi N$ collisions are taken as
in Refs. ~\cite{brat03,Cass01}.  Here we show only the total cross
sections for $D+\bar D$ productions in Fig. \ref{xs_pp_pip}.

In order to study the effect of charmonium rescattering we adopt the
following dissociation cross sections of charmonia with baryons
independent of the energy (in line with the recent NA50 and NA60
compilations~\cite{NA60,NA50pA}):
\begin{eqnarray}
&& \sigma_{c\bar{c}B} = 4.18 \ {\rm mb}; \label{sigmacB} \\
&&\sigma_{J/\Psi B} = 4.18 \ {\rm mb}; \ \sigma_{\chi_c B} = 4.18 \
{\rm mb}; \  \sigma_{\Psi^\prime B} = 7.6 \ {\rm mb}.
\nonumber\end{eqnarray}
In (\ref{sigmacB}) the cross section $\sigma_{c\bar{c}B}$ stands for
a (color dipole)  pre-resonance ($c\bar{c})$ - baryon cross section,
since the $c\bar{c}$ pair produced initially cannot be identified
with a particular hadron due to the uncertainty relation in energy
and time. For the life-time of the pre-resonance $c\bar{c}$ pair (in
it's rest frame) a value of $\tau_{c\bar{c}}$ = 0.3 fm/c is assumed
following Ref. ~\cite{Kharz}. This value corresponds to the mass
difference of the $\Psi^\prime$ and $J/\Psi$.

For $D, D^*, \bar{D}, \bar{D}^*$ - meson ($\pi, \eta, \rho,
\omega$) scattering we address to the calculations from Ref.
\cite{Konew,Ko} which predict elastic cross sections in the range
of 10--20 mb depending on the size of the formfactor employed. As
a guideline we use a constant cross section of 10 mb for elastic
scattering with mesons and also baryons, although the latter might
be even higher for very low relative momenta. Since the $D$-meson
dynamics is of minor importance for charmonium regeneration at SPS
energies we discard a more detailed description.

\subsection{The comover absorption model}

As already pointed out before, the $J/\Psi, \chi_c, \Psi^\prime$
formation cross sections by open charm mesons or the inverse comover
dissociation cross sections are not well known and the significance
of these channels is discussed controversely in the
literature~\cite{Bernd,BMS,Rafelski,Redlich,I2,I3,KoO}. We here
follow the concept of Refs.~\cite{brat03,brat04} and introduce a
simple 2-body transition model with a single parameter $|M_0|^2$,
that allows to implement the backward reactions uniquely by
employing detailed balance for each individual channel.

Since the charmonium-meson dissociation and backward reactions typically
occur with low relative momenta (`comovers') it is legitimate to
write the cross section for the process $1+2\to 3+4$ as
\begin{equation}
\label{model}
 \sigma_{1+2\to 3+4}(s) = 2^4 \frac{E_1 E_2 E_3 E_4}{s}
|\tilde M_i|^2 \left(\frac{m_3+m_4}{\sqrt{s}}\right)^6  \frac{p_f}{p_i},
\end{equation}
 where $E_k$ denotes the energy of hadron $k$
$(k=1,2,3,4)$, respectively. The initial and final momenta for fixed
invariant energy  $\sqrt{s}$ are given by
\begin{eqnarray}
p_i^2 = \frac{(s-(m_1+m_2)^2)(s-(m_1-m_2)^2)}{4s}, \nonumber\\
p_f^2 = \frac{(s-(m_3+m_4)^2)(s-(m_3-m_4)^2)}{4s},
\label{moment}
\end{eqnarray}
where $m_k$ denotes the mass of hadron $k$. In (\ref{model})
$|\tilde M_i|^2$ ($i=\chi_c, J/\psi, \psi^\prime$) stands for the
effective matrix element squared, which for the different 2-body
channels is taken of the form
\begin{eqnarray}
&&\hspace*{-3mm}|\tilde M_i|^2 =|M_i|^2  \ \ {\rm for} \
    \ (\pi,\rho)+(c\bar c)_i \to D+\bar{D} \label{mod}\\
&&\hspace*{-3mm}|\tilde M_i|^2 = 3 |M_i|^2  \ \ {\rm for} \
    \ (\pi,\rho)+(c\bar c)_i  \to D^*+\bar{D}, \ D+\bar{D}^*, \ D^* + \bar{D}^* \nonumber\\
&&\hspace*{-3mm}|\tilde M_i|^2 = \frac{1}{3} |M_i|^2 \ \ {\rm for}\
    \ (K,K^*)+(c\bar c)_i  \to D_s + \bar{D}, \ \bar{D}_s + D \nonumber \\
&&\hspace*{-3mm}|\tilde M_i|^2 =  |M_i|^2  \ \ {\rm for} \
    \ (K,K^*)+(c\bar c)_i  \to D_s + \bar{D}^*, \ \bar{D}_s + D^*,\ D^*_s + \bar{D}, \nonumber \\
&&\phantom{|\tilde M_i|^2 =  |M_i|^2  \ \ {\rm for} \ \
(K,K^*)+(c\bar c)_i  \to D_s + \bar{D}^*, \ }     \bar{D}^*_s + D, \
\bar{D}^*_s + D^* \nonumber
\end{eqnarray}
The relative factors of 3 in (\ref{mod}) are guided by the sum
rule studies in~\cite{korean} which suggest that the cross section
is increased whenever a vector meson $D^*$ or $\bar{D}^*$ appears in
the final channel while another factor of 1/3 is introduced for each
$s$ or $\bar{s}$ quark involved. The factor $\left(
{(m_3+m_4)}/{\sqrt{s}} \right)^6 $ in (\ref{model}) accounts for the
suppression of binary channels with increasing $\sqrt{s}$ and has
been fitted to the experimental data for the reactions $\pi + N
\rightarrow \rho+N, \omega+N, \phi+N, K^+ +\Lambda$ in Ref.
\cite{CaKo}.

We use (for simplicity) the same matrix elements for the
dissociation of all charmonium states $i$ ($i=\chi_c, J/\psi,
\psi^\prime$) with mesons:
\begin{eqnarray}
 |M_{J/\Psi}|^2 = |M_{\chi_c}|^2 = |M_{\Psi^\prime}|^2 = |M_0|^2.
\label{MatrElem}
\end{eqnarray}
We note that in Ref.~\cite{brat03} the parameter $|M_0|^2$ was fixed
by comparison to the $J/\Psi$ suppression data from the NA38 and
NA50 Collaborations for S+U and Pb+Pb collisions at 200 and 158
A$\cdot$GeV, respectively. In the present study, however, this
parameter has to be readjusted in accordance with the updated value
of the cross section~(\ref{sigmacB}) of charmonium dissociation on
baryons (following the latest NA50 and NA60
analysis~\cite{NA60,NA50pA}). The best fit is obtained for
$|M_0|^2=0.18$~fm$^2$/GeV$^2$.

The advantage of the model introduced in \cite{brat03,brat04} is
that detailed balance for the binary reactions can be employed
strictly for each individual channel, {\it i.e.}
\begin{eqnarray}
\!\!\sigma_{3+4 \rightarrow 1+2}(s) =
 \sigma_{1+2 \rightarrow 3+4}(s)
\frac{(2S_1+1)(2S_2+1)}{(2S_3+1)(2S_4+1)} \ \frac{p_i^2}{p_f^2}, \
\label{balance}
\end{eqnarray}
 and the role of the backward reactions
($(c\bar c)_i$+meson formation by $D+\bar{D}$ flavor exchange) can
be explored without introducing any additional parameter once
$|M_i|^2$ is fixed. In Eq. (\ref{balance}) the quantities $S_j$
denote the spins of the particles, while ${p_i^2}$ and ${p_f^2}$
denote the cms momentum squared in the initial and final channels,
respectively. The uncertainty in the cross sections (\ref{balance})
is of the same order of magnitude as that in Lagrangian approaches
using {\it e.g.} $SU(4)_{flavor}$ symmetry \cite{Konew,Ko}, since
the formfactors at the vertices are essentially
unknown~\cite{korean}. It should be pointed out that the comover
dissociation channels for charmonia are described in HSD with the
proper individual thresholds for each channel in contrast to the
more schematic comover absorption model~\cite{Capella}.

We recall that (as in Refs.
\cite{brat03,Cass01,Geiss99,Cass97,CassKo}) the charm degrees of
freedom in the HSD approach are treated perturbatively and that
initial hard processes (such as $c\bar{c}$ or Drell-Yan production
from $NN$ collisions) are `precalculated' to achieve a scaling of
the inclusive cross section with the number of projectile and target
nucleons as $A_P \times A_T$ when integrating over impact parameter.
For fixed impact parameter $b$ the $c\bar{c}$ yield then scales with
the number of binary hard collisions $N_{bin}$ ({\it cf.} Fig. 8 in
Ref.~\cite{Cass01}).

\subsection{Implementation of the `threshold scenario'}

The HSD transport model allows to calculate the energy-momentum
tensor $T_{\mu \nu}(x)$ for all space-time points $x$ and thus the
energy density $\varepsilon(x)$ in the local rest frame. In order to
exclude contributions to $T_{\mu \nu}$ from noninteracting
nucleons in the intial phase all nucleons without prior
interactions are discarded in the rapidity intervals
$[y_{tar}-0.4, y_{tar}+0.4]$ and $[y_{pro}-0.4, y_{pro}+0.4]$
where $y_{tar}$ and $y_{pro}$ denote projectile and target
rapidity, respectively. Note that the initial rapidity
distributions of projectile and target nucleons are smeared out by
about $\pm 0.4$ due to Fermi motion.

In the actual calculation the initial grid has a dimension of 1 fm
$\times$ 1 fm $\times$ 1/$\gamma_{cm}$ fm, where $\gamma_{cm}$
denotes the Lorentz $\gamma$-factor in the nucleon-nucleon
center-of-mass system. After the time of maximum overlap $t_m$ of
the nuclei the grid-size in beam direction $\Delta z_0 =
1/\gamma_{cm}$ [fm] is increased linearly in time as $\Delta z =
\Delta z_0 + a (t-t_m)$, where the parameter $a$ is chosen in a way
to keep the particle number in the local cells roughly constant
during the longitudinal expansion of the system. In this way local
fluctuations of the energy density $\varepsilon(x)$ due to
fluctuations in the particle number are kept low.

\begin{figure}[!]
\centerline{\psfig{figure=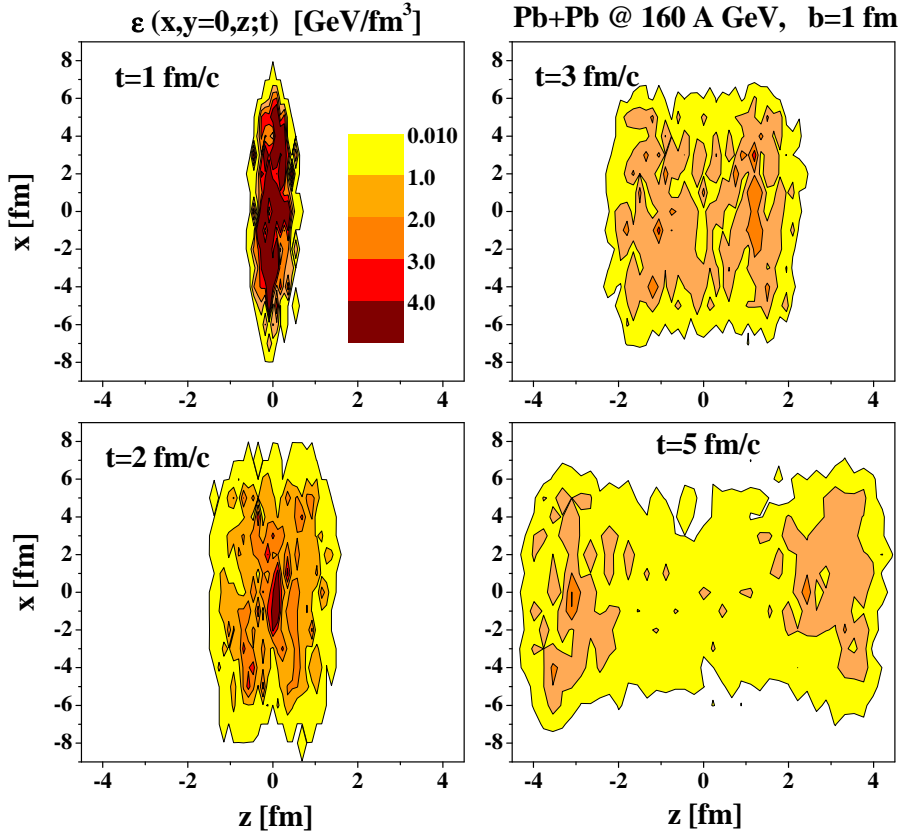,width=0.85\textwidth}}
\caption{The energy density $\varepsilon(x,y=0,z;t)$ from HSD for a
Pb+Pb collision at 160 A$\cdot$GeV and impact parameter $b=1$ fm in
terms of contour lines (0.01, 1, 2, 3, 4 GeV/fm$^3$) for times of 1,
2, 3 and 5 fm/c (from contact). Note that noninteracting nucleons
have been discarded in the actual calculation of the energy-momentum
tensor.} \label{F1}
\end{figure}

As an example we display in Fig. \ref{F1} the energy density
$\varepsilon(x,y=0,z;t)$ for a Pb+Pb collision at 160 A$\cdot$GeV and
impact parameter $b=1$ fm in terms of contour lines for times of
1, 2, 3 and 5 fm/c (from contact). It is clearly seen that energy
densities above 4 GeV/fm$^3$ are reached in the early overlap
phase of the reaction and that $\varepsilon(x)$ drops within a few
fm/c below 1 GeV/fm$^3$ in the center of the grid. On the other
hand the energy density in the region of the leading particles -
moving almost with the velocity of light - stays above 1
GeV/fm$^3$ due to Lorentz time dilatation since the time $t$ here
is measured in the nucleon-nucleon center-of-mass system. Note
that in the local rest frame of the leading particles the
eigentime $\tau$ is roughly given $\tau \approx t/\gamma_{cm}$
with $\gamma_{cm} \approx 9.3$.

\begin{figure}[!]
\centerline{\psfig{figure=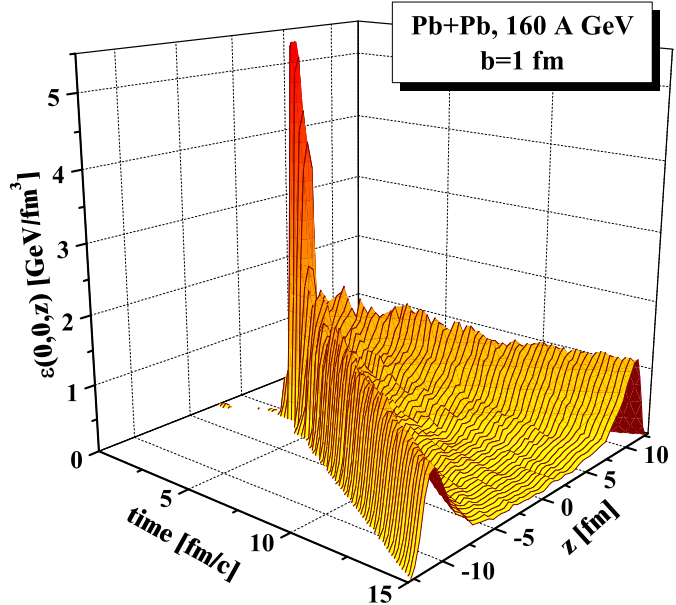,width=0.95\textwidth}}
\caption{The energy density $\varepsilon(x=0,y=0,z;t)$ from HSD for a
Pb+Pb collision at 160 A$\cdot$GeV and impact parameter $b=1$ fm on
a linear scale. Note that noninteracting nucleons have been
discarded in the actual calculation of the energy-momentum tensor
such that $\varepsilon(x)\ne 0$ only after contact of the two Pb nuclei
which is $\sim$ 2 fm/c.} \label{F2}
\end{figure}

Another view of the space time evolution of the energy density is
given in Fig.~\ref{F2} where we display $\varepsilon(x=0,y=0,z;t)$ for
the same system as in Fig. \ref{F1} on a linear scale. The contact
time of the two Pb nuclei here is 2 fm/c and the overlap phase of
the Lorentz contracted nuclei is identified by a sharp peak in
space-time which is essentially given by the diameter of the nuclei
divided by $\gamma_{cm}$. As noted before, the energy density in the
center of the reaction volume ($z \approx$ 0) drops fast below 1
GeV/fm$^3$ whereas the ridges close to the lightcone basically stem
from the leading ends of the strings formed in the early
nucleon-nucleon collisions. In these space-time regions all reaction
rates are reduced by the factor $\sim 1/\gamma_{cm}$ such that the
transport calculations have to be carried to large times of several
hundred fm/c in order to catch the dynamics and decays in these
regions. In the central regime, however, all interaction rates
vanish after about 15 fm/c. Since the $c,\bar{c}$ pairs are produced
dominantly at midrapidity with a small spread in rapidity ($\sigma_y
\approx 0.8$ at 160 A$\cdot$GeV) it is the central region that is of
primary interest for this study.

\begin{figure*}[!]
{\psfig{figure=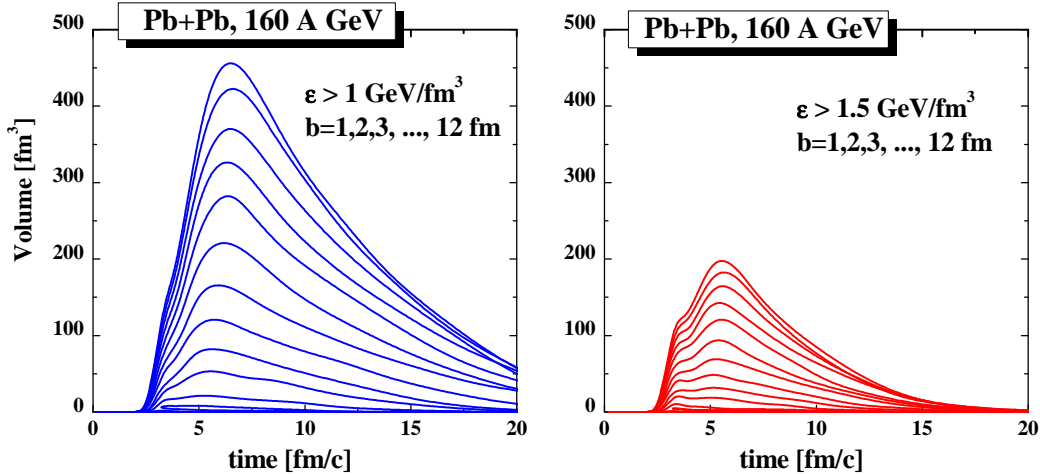,width=0.75\textwidth}} \caption{The volume
$V(\varepsilon_c;t)$ (\protect\ref{volume})  from HSD for Pb+Pb
collisions at 160 A$\cdot$GeV and impact parameter $b=1,2,..,12$ fm
for $\varepsilon > \varepsilon_c$ = 1 GeV/fm$^3$ (left part) and
$\varepsilon > \varepsilon_c$ = 1.5 GeV/fm$^3$ (right part).}
\label{F3}
\end{figure*}

Of further interest is the size of the total volume (measured in the
nucleon-nucleon cms) with an energy density above a certain cut
$\varepsilon_c$, {\it i.e.}
\begin{equation} \label{volume}
V(\varepsilon_c;t): = \int d^3r \ \Theta(\varepsilon({\bf r};t) -
\varepsilon_c),
\end{equation}
which quantifies the volume for charmonium dissolution as a function
of time $t$. The corresponding information is displayed in Fig.
\ref{F3} for $\varepsilon_c$ = 1 GeV/fm$^3$ (left part) and 1.5
GeV/fm$^3$ (right part) as a function of time for impact parameter
b=1 to 12 fm (in steps of $\Delta$ b = 1 fm). These volumina may be
compared to the Lorentz contracted eigenvolume of a Pb nucleus that
is about 160 fm$^3$ in the cms. It is clearly seen that hadron
formation and the explosion (or expansion) of the system lead to
larger volumina $V(\varepsilon_c;t)$ in central reactions especially
for $\varepsilon_c$ = 1 GeV/fm$^3$. Note, however, that charmonia
dynamically cannot explore the whole volume displayed in
Fig.~\ref{F3}, since this volume is dominated by the space-time
regimes close to the lightcone ({\it cf.} Fig.~\ref{F2}), where
practically no charmonia appear at 160 A$\cdot$GeV.

The `threshold scenario' for charmonium dissociation now is
implemented in a straight forward way: whenever the local energy
density $\varepsilon(x)$ is above a threshold value $\varepsilon_j$, where
the index $j$ stands for $J/\Psi, \chi_c, \Psi^\prime$, the
charmonium is fully dissociated to $c + \bar{c}$. The default
threshold energy densities adopted are $\varepsilon_1 = 16$ GeV/fm$^3$
for $J/\Psi$, $\varepsilon_2 = 2$ GeV/fm$^3$ for $\chi_c$, and
$\varepsilon_3 =2 $ GeV/fm$^3$ for $ \Psi^\prime$. The reformation of
charmonia at the phase boundary to the hadronic system is discarded
in view of the very low charm quark density at SPS energies.

\section{Comparison to data}

We directly step on with results for the charmonium suppression at
SPS energies in comparison with the experimental data from the NA50
and NA60 Collaborations.  These Collaborations present their results
on $J/\Psi$ suppression as the ratio of the dimuon decay of $J/\Psi$
relative to the Drell-Yan background from 2.9 - 4.5 GeV invariant
mass as a function of the transverse energy $E_T$, or alternative,
as a function of the number of participants $N_{{\rm part}}$, {\it
i.e.}
\begin{equation} \label{rat} B_{\mu\mu}\sigma(J/\Psi) /
\sigma(DY)|_{2.9-4.5},
\end{equation}
where $B_{\mu\mu}$ is the branching ratio for $J/\Psi\to
\mu^+\mu^-$.

\begin{figure*}[!]
\centerline{\psfig{figure=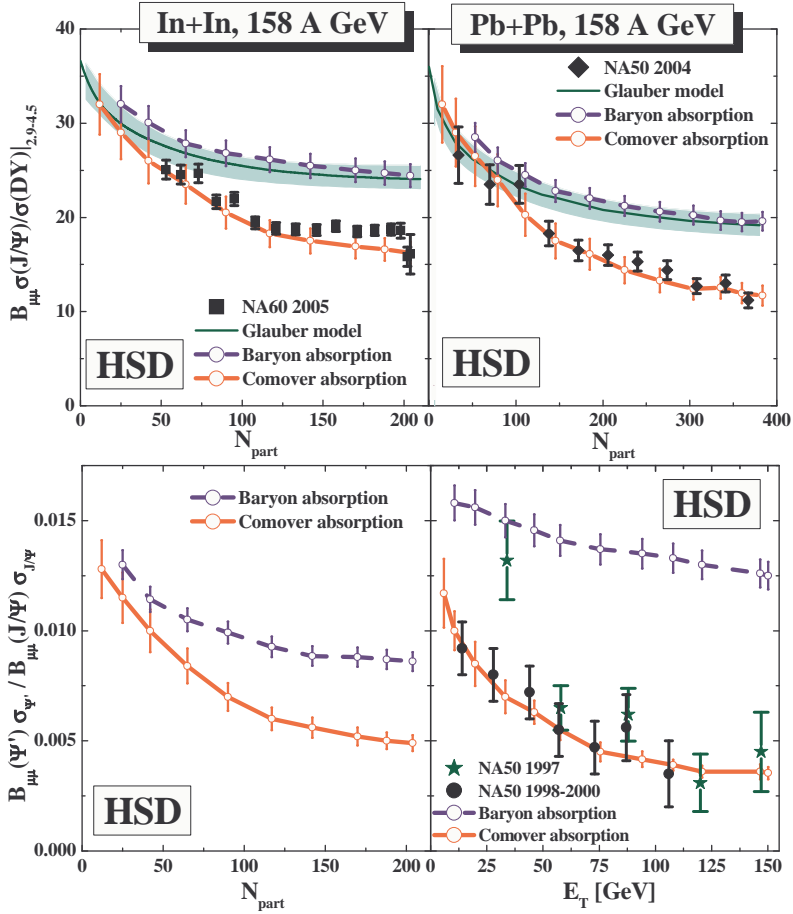,width=0.75\textwidth}}
\caption{The ratio $B_{\mu\mu}\sigma(J/\Psi) / \sigma(DY)$  as a
function of the number of participants in In+In (l.h.s.) and Pb+Pb
reactions (r.h.s.) at 158 A$\cdot$GeV. The full symbols denote the
data from the NA50 and NA60 Collaborations (from Refs.
\protect\cite{NA50O,NA60,NA50PsiPrime}), while the dashed (blue)
lines represent the HSD calculations including only dissociation
channels with nucleons. The lower parts of the figure show the HSD
results in the same limit for the $\Psi^\prime$ to $J/\Psi$ ratio as
a function of $N_{part}$ (for In+In) or the transverse energy $E_T$
(for Pb+Pb). The solid (red) lines show the HSD results for the
comover absorption model with a matrix element squared $|M_0|^2$ =
0.18 fm$^2$/GeV$^2$.  The (light blue) bands in the upper parts of
the figure give the estimate for the normal
 nuclear $J/\Psi$ absorption as calculated by the NA60 Collaboration.
 The vertical lines on
the graphs reflect the theoretical uncertainty due to limited
statistics of the calculations.} \label{figure1}
\end{figure*}

In the theoretical approaches we calculate the $J/\Psi$ survival
probability $S_{J/\Psi}$ defined as
\begin{equation} \label{supp} S_{J/\Psi} =
\frac{N^{J/\Psi}_{fin}}{N^{J/\Psi}_{BB}},
\end{equation}
where $ N^{J/\Psi}_{fin}$ and $N^{J/\Psi}_{BB}$ denote the final
number of $J/\Psi$ mesons and the number of $J/\Psi$'s produced
initially by $BB$ reactions, respectively. In order to compare our
calculated results to experimental data we need an extra input, i.e.
the normalization factor $B_{\mu\mu}\sigma_{NN}(J/\Psi) /
\sigma_{NN}(DY)$, which defines the $J/\Psi$ over Drell-Yan ratio
for elementary nucleon-nucleon collisions. We choose
$B_{\mu\mu}\sigma_{NN}(J/\Psi) / \sigma_{NN}(DY) = 36$ in line with
the NA60 compilation~\cite{NA60}.

Furthermore, the $\Psi^\prime$ suppression is presented
experimentally by the ratio
\begin{equation} \label{pis}
\frac{B_{\mu\mu}(\Psi^\prime\to \mu\mu)\sigma(\Psi^\prime)/\sigma(DY) }
{B_{\mu\mu}(J/\Psi\to \mu\mu)\sigma(J/\Psi) / \sigma(DY)}.
\end{equation}
In our calculations we adopt this ratio to be 0.0165 for
nucleon-nucleon collisions, which is again based on the average over
$pp, pd, pA$ reactions~\cite{NA50_03}.

\begin{figure*}[!]
\centerline{\psfig{figure=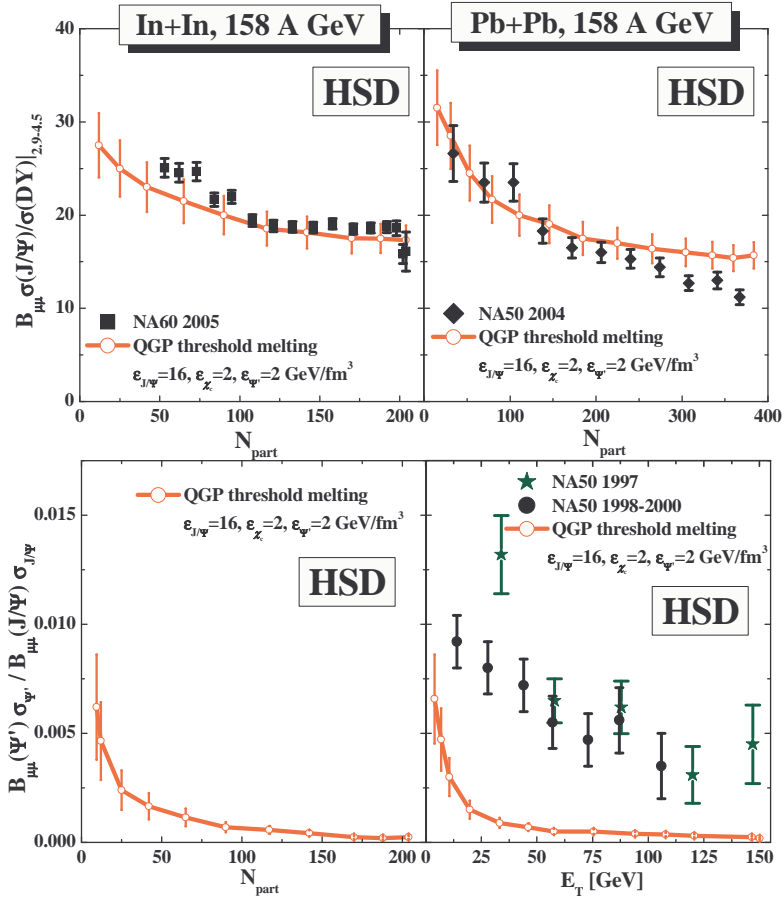,width=0.75\textwidth}}
\caption{Same as Fig. \protect\ref{figure1} but for the `QGP
threshold scenario' with $\varepsilon_{J/\Psi} = 16$ GeV/fm$^3$,
$\varepsilon_{\chi_c} = 2$ GeV/fm$^3$ = $\varepsilon_{\Psi^\prime}$
while discarding comover absorption, {\it i.e.} for $|M_0|^2 = 0$. }
\label{figure3}
\end{figure*}

In order to investigate the anomalous charmonium suppression in
nucleus-nucleus collisions we first show in Fig. \ref{figure1} the
calculated ratio $B_{\mu\mu}\sigma(J/\Psi) / \sigma(DY)$ as a
function of $N_{{\rm part}}$ for Pb+Pb and In+In collisions at 158
A$\cdot$GeV (upper plots) in the nuclear suppression scenario, i.e.
without comover dissociation or `QGP threshold suppression'. The
dashed (blue) lines stand for the HSD result  while the (light blue)
bands  give the estimate for the normal nuclear $J/\Psi$ absorption
as calculated by the NA60 Collaboration. The normal nuclear
suppression from HSD is seen to be slightly lower than the (model
dependent) estimate from NA60, however, agrees quite well with their
model calculations for more central reactions. The various
experimental data points have been taken from
Refs.~\cite{NA50O,NA60,NA50PsiPrime}. It is clearly seen that the
charmonium dissociation with only nucleons is insufficient to
describe the data for both systems.  In the lower part of Fig.
\ref{figure1} we compare the calculated ratio $\Psi^\prime$ over
$J/\Psi$ as a function of $N_{part}$ (for In+In reactions) or the
transverse energy $E_T$ (for Pb+Pb collisions), respectively, in
comparison to the data available. The Pb+Pb data demonstrate that
the centrality dependence as well as the absolute ratio cannot be
explained by nuclear dissociation channels alone which is a well
known fact in the community.

Apart from the statistical uncertainties in the calculations -
reflected by the vertical lines on the theoretical graphs in Fig.
\ref{figure1} - some dependence on the model parameters enters the
actual numbers in Fig. \ref{figure1}. The charmonium nuclear
absorption cross section is
considered to be 'fixed' by the NA50/NA60 compilations and we have
taken the same cross section for the 'pre-resonance' cross section
 for the $J/\Psi$ and $\chi_c$. Accordingly the life-time of the
pre-resonance state ($\tau_{c\bar{c}}$ = 0.3 fm/c) has no impact on
the absorption with baryons as far as the $J/\Psi$ and $\chi_c$
mesons are concerned. Only for $\Psi^\prime$ collisions with baryons
this plays a role since the $\Psi^\prime$ + baryon cross section is
larger (7.6 mb). Consequently the $J/\Psi$ suppression (including
the feed down from $\chi_c$) does not depend on $\tau_{c\bar{c}}$.
The finite life-time, however, plays a role for $\Psi^\prime$
suppression as can be seen in the lower right part of Fig.
\ref{figure1} since it leads to a larger baryon absorption of
$\Psi^\prime$ (relative to $J/\Psi$) with increasing centrality by
about 22\%. According to our understanding the life-time of 0.3 fm/c
is a lower limit (in line with the Heisenberg uncertainty relation);
it leads to a maximum suppression of $\Psi^\prime$ relative to
$J/\Psi$ with centrality. On the other hand, a very large life-time
$\tau_{c\bar{c}}$ will lead to a constant ratio of $\Psi^\prime$ to
$J/\Psi$ with centrality since only the pre-resonance cross section
will apply. Accordingly, the $\Psi^\prime$ to $J/\Psi$ ratio is
driven by the value of $\tau_{c\bar{c}}$ since the ratio of the
dissociation cross section for the formed mesons is fixed by the
ratio of their mean square radii. However, independently on the
life-time $\tau_{c\bar{c}}$ of the pre-resonance state the
experimental data will be badly missed for the $\Psi^\prime$ to
$J/\Psi$ ratio if only baryon dissociation is included, because the
case considered here already provides a maximum suppression of the
$\Psi^\prime$ for interactions with baryons.

As a next step we add the comover dissociation channels within the
model described in Section 2.1 for a matrix element squared
$|M_0|^2$ = 0.18 fm$^2$/GeV$^2$. Note that in this case the
charmonium reformation channels are incorporated, too, but could be
discarded since the charmonium regeneration is negligible at SPS
energies ({\it cf.} Ref.~\cite{brat03}). The extra suppression of
charmonia by comovers is seen in Fig. \ref{figure1} (solid (red)
lines) to match the $J/\Psi$ suppression in In+In and Pb+Pb as well
as the $\Psi^\prime$ to $J/\Psi$ ratio (for Pb+Pb) rather well. The
more recent data (1998-2000) for the $\Psi^\prime$ to $J/\Psi$ ratio
agree with the HSD prediction within error bars. This had been a
problem in the past when comparing to the 1997 data (dark green
stars). We conclude that the comover absorption model presently
cannot be ruled out on the basis of the available data sets within
error bars. The $\Psi^\prime$ to $J/\Psi$ ratio for In+In versus
centrality is not yet available from the experimental side but the
theoretical predictions are provided in Fig. \ref{figure1} and might
be approved/falsified in near future.

Some comments to the comover absorption model appear in place: As
shown in Fig. 7.2 of Ref. \cite{Cass99} the comover densities in
central Pb+Pb collisions at 158 A$\cdot$GeV become quite large and
almost reach 2/fm$^3$ in the maximum which appears high for `free'
mesons with an eigenvolume of about 1 fm$^3$. However, the
quasi-particle mesons considered here dynamically should not be
identified with `free' meson states that show a long polarization
tail in the vacuum. As known from lattice QCD the correlators for
pions and $\rho$-mesons survive well above the critical temperature
$T_c$, such that `dressed' mesons, i.e spectral densities with the
quantum numbers of the pseudo-scalar and vector (isovector) modes,
also show up at high energy density (similar to the $J/\Psi$
discussed above \cite{KarschJP,HatsudaJP,Karsch2}). Such `dressed'
mesons are expected to have a shorter polarization tail - since the
reference vacuum has changed and the vacuum polarization decreases -
and thus are
 much smaller in size. In any case, these `states'
(or resonances) will dissociate charmonia by a quark rearrangement
interaction in exchanging a light quark with a $c$ or $\bar{c}$ quark.

\begin{figure*}[!]
\centerline{\psfig{figure=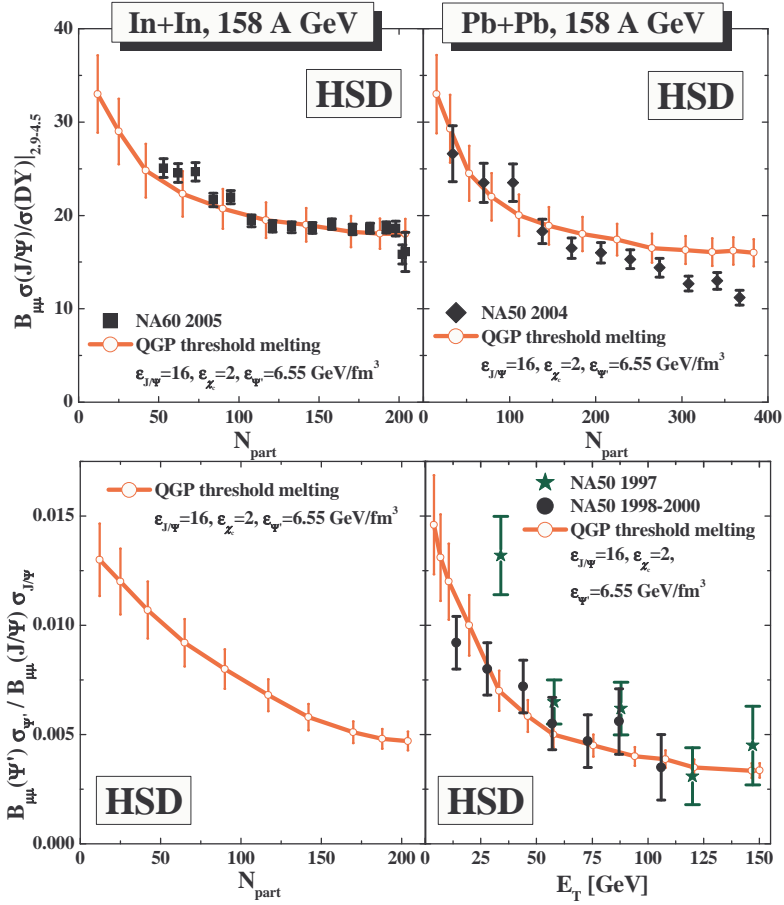,width=0.75\textwidth}}
\caption{Same as Fig. \protect\ref{figure1} but for the `QGP
threshold scenario' with $\varepsilon_{J/\Psi} = 16$ GeV/fm$^3$,
$\varepsilon_{\chi_c} = 2$ GeV/fm$^3$,  $\varepsilon_{\Psi^\prime}$
= 6.55 GeV/fm$^3$ while discarding comover absorption, {\it i.e.}
for $|M_0|^2 = 0$. } \label{figure4}
\end{figure*}

The results for the `threshold scenario' are displayed in
Fig.~\ref{figure3} in comparison to the same data for the thresholds
$\varepsilon_{J/\Psi} = 16$ GeV/fm$^3$, $\varepsilon_{\chi_c} = 2$
GeV/fm$^3$ = $\varepsilon_{\Psi^\prime}$ while discarding any
dissociation with comovers, i.e. $|M_0|^2$ =0.
In this scenario the
$J/\Psi$ suppression is well described for In+In but the suppression
is slightly too weak for very central Pb+Pb reactions. This result
emerges since practically all $\chi_c$ and $\Psi^\prime$ dissolve
for $N_{part} >$ 100 in both systems whereas the $J/\Psi$ itself
survives at the energy densities reached in the collision. Since the
nucleon dissociation is a flat function of $N_{part} $ for central
reactions the total absorption strength is flat, too. The deviations
seen in Fig. \ref{figure3} might indicate a partial melting of the
$J/\Psi$ for $N_{part} > $ 250, which is not in line with the
lattice QCD calculations claiming at least $\varepsilon_{J/\Psi} >$
5 GeV/fm$^3$. In fact, a lower threshold of  5 GeV/fm$^3$ (instead
of 16 GeV/fm$^3$) for the $J/\Psi$ has practically no effect on the
results shown in Fig. \ref{figure3}. Furthermore, a threshold energy
density of 2 GeV/fm$^3$ for the $\Psi^\prime$ leads to a dramatic
reduction of the $\Psi^\prime$ to $J/\Psi$ ratio which is in severe
conflict with the data (lower part of Fig.~\ref{figure3}). Also note
that due to energy density fluctuations in reactions with fixed
$N_{part} $ (or $E_T$) there is no step in the suppression of
$J/\Psi$ versus centrality as pointed out before by Gorenstein et
al. in Ref.~\cite{Goren}.

\begin{figure*}[!]
\centerline{\psfig{figure=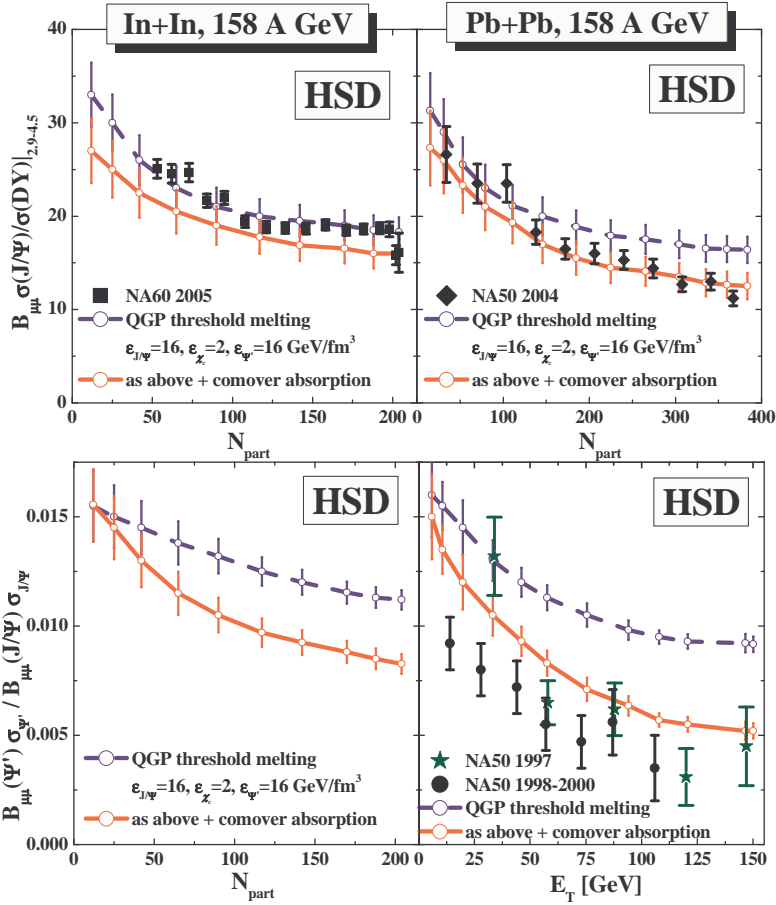,width=0.75\textwidth}}
\caption{Same as Fig. \protect\ref{figure1} but for the `QGP
threshold scenario' with $\varepsilon_{J/\Psi} = 16$ GeV/fm$^3$,
$\varepsilon_{\chi_c} = 2$ GeV/fm$^3$,  $\varepsilon_{\Psi^\prime}$ = 16
GeV/fm$^3$ (dashed blue lines) and with additional comover
absorption for $|M_0|^2 = 0.09$ fm$^2$/GeV$^2$ (red solid lines). }
\label{figure5}
\end{figure*}

Since in the `threshold scenario' the $J/\Psi$ suppression is rather
well accounted for by the melting of the $\chi_c$ it might be
tempting to `extract' a dissociation energy density for the
$\Psi^\prime$ via the $\Psi^\prime$ to $J/\Psi$ ratio. This may
indeed be achieved for $\varepsilon_{\Psi^\prime}$ = 6.55 GeV/fm$^3$
as shown in Fig. \ref{figure4} where now also the $\Psi^\prime$ to
$J/\Psi$ ratio is well reproduced for Pb+Pb. Respective predictions
for In+In within this scenario are presented in the lower left part
of Fig. \ref{figure4} and wait for confirmation or disproof. We
recall, however, that the energy density of 6.55 GeV/fm$^3$ for
$\Psi^\prime$ dissociation is not supported by present lattice
calculations such that this limit should be ruled out.

As the last model scenario we assume that both $J/\Psi$ and
$\Psi^\prime$ are practically not dissociated in nucleus-nucleus
reactions at SPS energies but let the $\chi_c$ melt above 2
GeV/fm$^3$ as before. In this case one may estimate the maximum
suppression by comovers that is compatible with the data. This
combined scenario allows to fix a matrix element squared $|M_0|^2
\approx$ 0.09  fm$^2$/GeV$^2$ that gives results still compatible with
the $J/\Psi$ suppression for In+In and Pb+Pb (solid lines in Fig.
\ref{figure5}). The dashed lines in Fig. \ref{figure5} show the
results without any comover absorption and are due to the melting
of the $\chi_c$ above 2 GeV/fm$^3$ alone. However, in this
combined approach the $\Psi^\prime$ to $J/\Psi$ ratio turns out to
be systematically too high in comparison to the data for Pb+Pb
such that this scenario should be ruled out, too, in particular
with respect to the extreme threshold for $\Psi^\prime$
dissociation.

\begin{figure*}[!]
\centerline{\psfig{figure=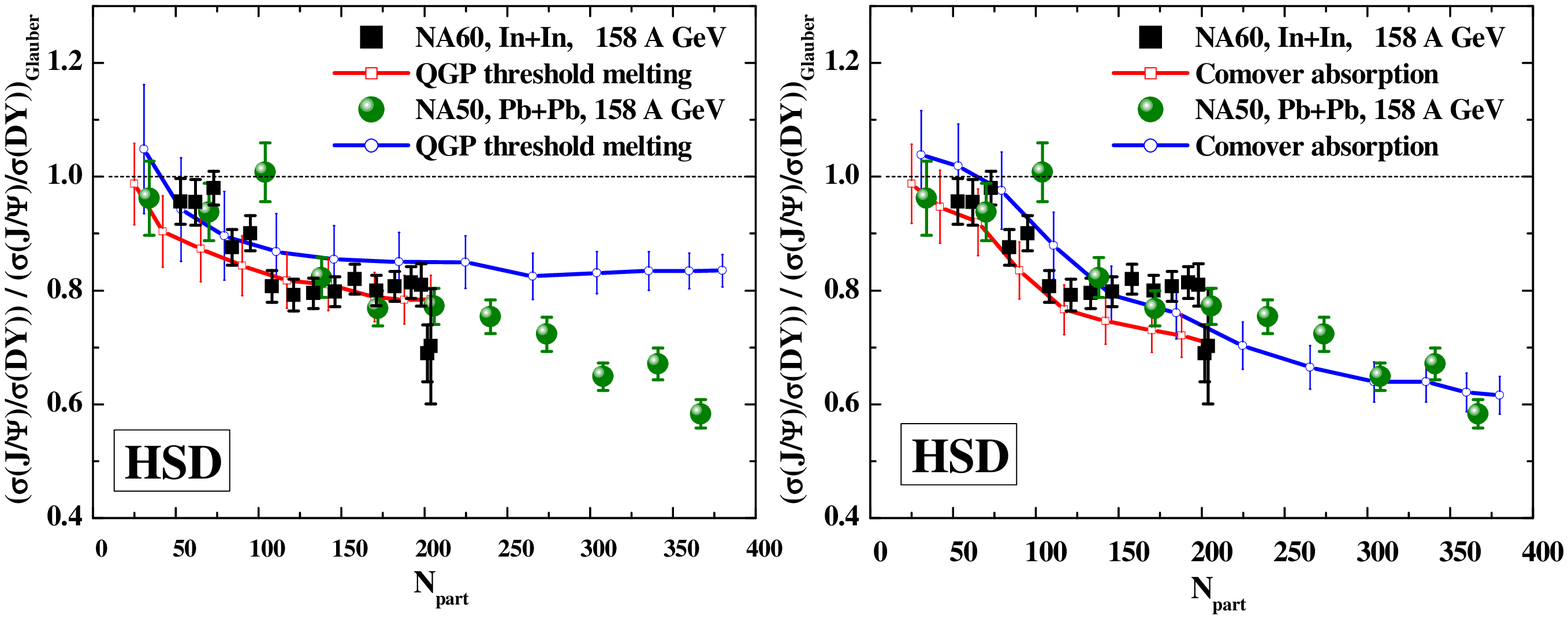,width=0.98\textwidth}}
\caption{The ratio $B_{\mu\mu}\sigma(J/\Psi) / \sigma(DY)$  as a
function of the number of participants $N_{part}$ in In+In (red line
with open squares) and Pb+Pb reactions (blue line with open circles)
at 158 A$\cdot$GeV relative to the normal nuclear absorption given
by the straight black line. The full dots and squares denote the
respective data from the NA50 and NA60 Collaborations. The model
calculations reflect the comover absorption model (right part) and
the `QGP threshold scenario' (left part) with $\varepsilon_{J/\Psi}
= 16$ GeV/fm$^3$, $\varepsilon_{\chi_c} = 2$ GeV/fm$^3$,
$\varepsilon_{\Psi^\prime}$ = 6.55 GeV/fm$^3$ while discarding
comover absorption.
 } \label{figure6}
\end{figure*}

Since the NA60 Collaboration prefers to represent their data in a
model dependent way by plotting their experimental results relative
to the normal nuclear absorption model we additionally show in Fig.
\ref{figure6} our calculations for In+In (red lines with open
squares) and Pb+Pb (blue lines with open circles) as a function of
the number of participants $N_{part}$ relative to the normal nuclear
absorption given by the straight black line (according to the NA60
compilation). The full dots and squares denote the respective data
from the NA50 and NA60 Collaborations. The model calculations
reflect the comover absorption model (right part) and the `QGP
threshold scenario' (left part) with $\varepsilon_{J/\Psi} =
16$~GeV/fm$^3$, $\varepsilon_{\chi_c} = 2$~GeV/fm$^3$,
$\varepsilon_{\Psi^\prime} = 6.55$~GeV/fm$^3$. Since only the
representation is different the message stays the same: The comover
absorption model follows slightly better the fall of the $J/\Psi$
survival probability with increasing centrality whereas the
`threshold scenario' leads to an approximate plateau in both
reactions for high centrality.

\section{Predictions for FAIR energies}

\begin{figure*}[!]
\centerline{\psfig{figure=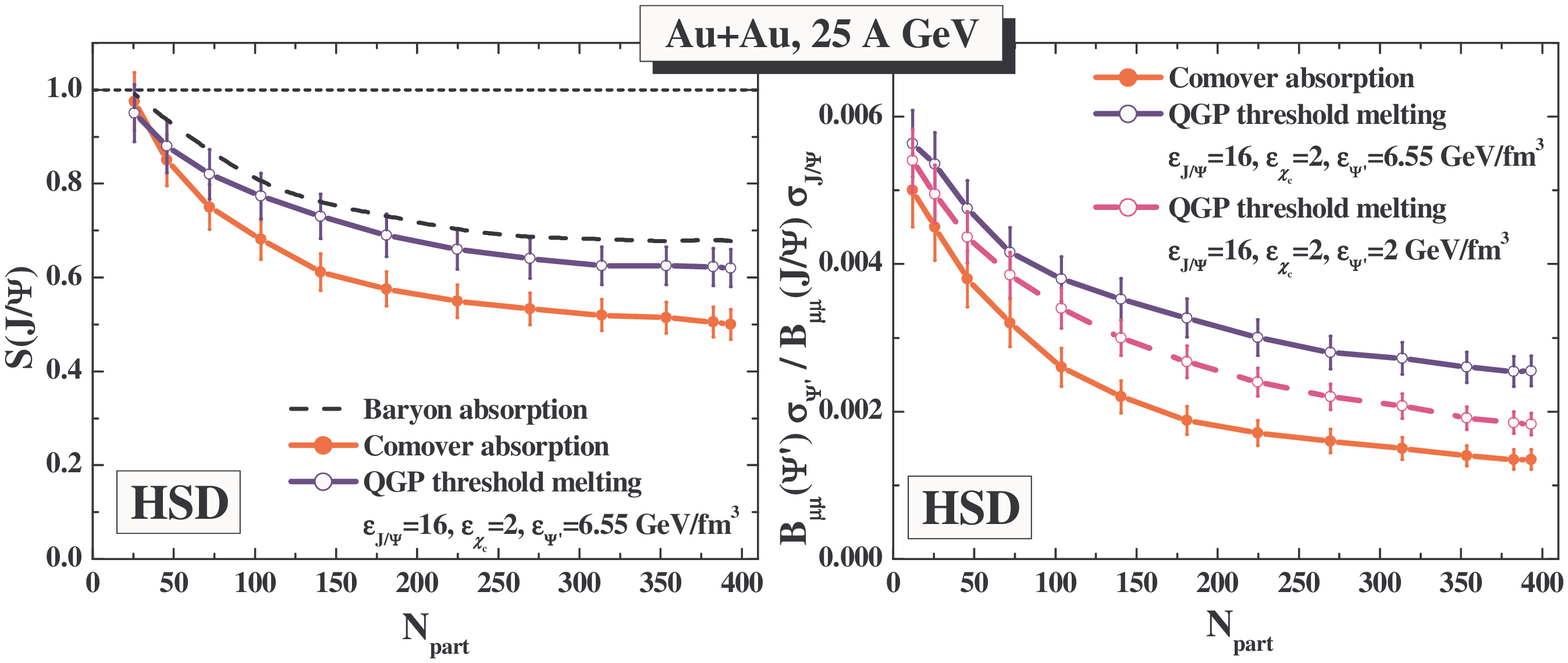,width=0.98\textwidth}}
\caption{The survival probability  $S_{J/\Psi}$ (left plot) and
ratio $\Psi^\prime$ to $J/\Psi$  (right plot) as a function of the
number of participants $N_{part}$ in Au+Au reactions at 25
A$\cdot$GeV.  The blue lines (with open dots) reflect the `threshold
scenario' for $\varepsilon_{J/\Psi} = 16$ GeV/fm$^3$,
$\varepsilon_{\chi_c} = 2$ GeV/fm$^3$, $\varepsilon_{\Psi^\prime}$ =
6.55 GeV/fm$^3$ while the violet line (the lower line with open dots
on the r.h.s.) stands for the 'threshold scenario' for
$\varepsilon_{J/\Psi} = 16$ GeV/fm$^3$, $\varepsilon_{\chi_c} = 2$
GeV/fm$^3$, $\varepsilon_{\Psi^\prime}$ = 2 GeV/fm$^3$. The solid
red lines (full dots) denote the results for the comover absorption
model with the standard matrix element squared $|M_0|^2$ = 0.18
fm$^2$/GeV$^2$. The dashed line (l.h.s.) represents the HSD
calculations including only dissociation channels with nucleons.}
\label{figure8}
\end{figure*}

The CBM Collaboration at GSI is aiming at charmonium measurements at
the future FAIR facility \cite{CBMprop}. This opens up the
possibility to explore the charmonium suppression mechanism at lower
bombarding energies of about 25 A$\cdot$GeV in Au+Au collisions.
First predictions for central reactions within the comover model
have been reported in Ref. \cite{Cass00}. Here we extend the earlier
studies to the full centrality dependence of the $J/\Psi$
suppression also within the `threshold scenario' and additionally
provide predictions for the $\Psi^\prime$ to $J/\Psi$ ratio. The
corresponding HSD results are displayed in Fig. \ref{figure8} for
the survival probability  $S_{J/\Psi}$ (left plot) and ratio
$\Psi^\prime$ to $J/\Psi$  (right plot) as a function of the number
of participants $N_{part}$. The blue lines reflect the `threshold
scenario' for $\varepsilon_{J/\Psi} = 16$ GeV/fm$^3$,
$\varepsilon_{\chi_c} = 2$ GeV/fm$^3$, $\varepsilon_{\Psi^\prime}$ =
6.55 GeV/fm$^3$ while the violet line stands for the `threshold
scenario' with a more realistic value of
$\varepsilon_{\Psi^\prime}$, i.e. $\varepsilon_{J/\Psi} = 16$
GeV/fm$^3$, $\varepsilon_{\chi_c} = 2$ GeV/fm$^3$,
$\varepsilon_{\Psi^\prime}$ = 2 GeV/fm$^3$. The solid red lines
denote the results for the comover absorption model with the
standard matrix element squared $|M_0|^2$ = 0.18 fm$^2$/GeV$^2$.

We note that in Au+Au reactions at 25 A$\cdot$GeV the standard
nuclear suppression of $J/\Psi$ (dashed line in the left plot)
almost coincides with the `threshold scenario' (solid line with open
dots in the left plot) since only a very low ammount of $\chi_c$ and
no $J/\Psi$ are melted at the energy densities reached in these
reactions. On the other hand the comover density decreases only
moderately when stepping down in energy from 158 A$\cdot$GeV to 25
A$\cdot$GeV such that the $J/\Psi$ survival probability in the
comover absorption model (lower solid line in the left part) is
substantially lower. This also holds for the $\Psi^\prime$ to
$J/\Psi$ ratio versus centrality where even a lower threshold
$\varepsilon_{\Psi^\prime}$ = 2 GeV/fm$^3$ leads to a ratio (middle
line in the right part) that is clearly above the result achieved in
the comover absorption model (lower line in the right part).
Consequently the different dissociation scenarios may well be
distinguished in future charmonium measurements at FAIR.

\section{Summary}

In summarizing this work we have found that present data on
charmonium suppression for Pb+Pb and In+In reactions at top SPS
energies compare well with microscopic transport calculations in the
comover model involving only a single parameter for the average
matrix element squared $|M_0|^2$ that fixes the strength of the
charmonium cross sections with comovers. This holds for the $J/\Psi$
suppression as well as the $\Psi^\prime$ to $J/\Psi$ ratio versus
collision centrality. The bare `QGP threshold scenario' gives
satisfying results for the $J/\Psi$ suppression for both systems at
158 A$\cdot$GeV but fails in the $\Psi^\prime$ to $J/\Psi$ ratio
since too many $\Psi^\prime$ already melt away for a critical energy
density of 2 GeV/fm$^3$ at 158 A$\cdot$GeV. Only when assuming the
$\Psi^\prime$ to dissolve above $\sim 6.5$ GeV/fm$^3$ a reasonable
description of all data is achieved in the `QGP threshold scenario';
this threshold, however, is not in accordance with present lattice
QCD calculations such that the `threshold scenario' meets severe
problems. This also holds for a combined model including `threshold
melting' and reduced `comover absorption' as shown in Section 3.

On the other hand the different scenarios can clearly be
distinguished at  FAIR energies (of about 25 A$\cdot$GeV) where the
centrality dependence of the $J/\Psi$ survival probability and the
$\Psi^\prime$ to $J/\Psi$ ratio are significantly lower in the
comover absorption model. This result comes about since the average
comover density decreases only moderately with lower bombarding
energy whereas the region in space-time with energy densities above
critical values of {\it e.g.} 2 GeV/fm$^3$ decreases rapidly and
ceases to exist below about 20 A$\cdot$GeV even in central
collisions. This ideally might open up the possibility to measure
excitation functions of the $J/\Psi$ survival probability and the
$\Psi^\prime$ to $J/\Psi$ ratio in central Au+Au collisions, where
clear steps would indicate the presence of `melting thresholds'
whereas a smooth excitation function would be in favor of the
comover absorption approach.

\section*{Acknowledgement}

The authors acknowledge inspiring discussions with A. Kostyuk,
E. Scomparin and P. Senger.


\end{document}